\documentclass[aps, prd, letterpaper, twocolumn, amsmath, amssymb, superscriptaddress, floatfix, nofootinbib, longbibliography]{revtex4-1}
\usepackage[T1]{fontenc}
\usepackage[english]{babel}
\usepackage[utf8]{inputenc}
\usepackage{graphicx}
\usepackage{color}
\usepackage{xcolor}
\usepackage{mathrsfs}
\usepackage{amsfonts}
\usepackage[colorlinks=true, linkcolor=blue, urlcolor=blue, citecolor=blue, anchorcolor=blue]{hyperref}
\usepackage{slashed}

\newcommand{\Ibb}{\ensuremath{\mathbb I} }
\newcommand{\Rbb}{\ensuremath{\mathbb R} }
\newcommand{\Tbb}{\ensuremath{\mathbb T} }
\newcommand{\cD}{\ensuremath{\mathcal D} }
\newcommand{\cDbar}{\ensuremath{\overline{\mathcal D}} }
\newcommand{\cF}{\ensuremath{\mathcal F} }
\newcommand{\cFbar}{\ensuremath{\overline{\mathcal F}} }
\newcommand{\vn}{\mathbf{n}}
\newcommand{\cN}{\ensuremath{\mathcal N} }
\newcommand{\cO}{\ensuremath{\mathcal O} }
\newcommand{\cQ}{\ensuremath{\mathcal Q} }
\newcommand{\cU}{\ensuremath{\mathcal U} }
\newcommand{\cUbar}{\ensuremath{\overline{\mathcal U}} }
\newcommand{\al}{\ensuremath{\alpha} }
\newcommand{\be}{\ensuremath{\beta} }

\newcommand{\Ga}{\ensuremath{\Gamma} }
\newcommand{\eps}{\ensuremath{\epsilon} }
\newcommand{\ka}{\ensuremath{\kappa} }
\newcommand{\la}{\ensuremath{\lambda} }
\newcommand{\lalat}{\ensuremath{\la_{\text{lat}}} }
\newcommand{\hatbmu}{\widehat{\boldsymbol{\mu}}}
\newcommand{\X}{\ensuremath{\!\times\!} }
\newcommand{\gsim}{\ensuremath{\gtrsim} }
\newcommand{\lsim}{\ensuremath{\lesssim} }
\newcommand{\vev}[1]{\ensuremath{\left\langle #1 \right\rangle} }
\newcommand{\pf}{\ensuremath{\text{pf}\,} }
\newcommand{\Tr}[1]{\ensuremath{\text{Tr}\left[ #1 \right]} }
\newcommand{\eq}[1]{Eq.~(\ref{#1})}
\newcommand{\fig}[1]{Fig.~\ref{#1}}
\newcommand{\refcite}[1]{Ref.~\cite{#1}}
\newcommand{\secref}[1]{Section~\ref{#1}}

\begin{document}
\title{Three-dimensional super-Yang--Mills theory on the lattice and dual black branes}

\author{Simon Catterall}
\email{smcatter@syr.edu}
\affiliation{Department of Physics, Syracuse University, Syracuse, New York 13244, United States}

\author{Joel Giedt}
\email{giedtj@rpi.edu}
\affiliation{Department of Physics, Applied Physics and Astronomy, Rensselaer Polytechnic Institute, 110 8th Street, Troy, New York 12065, United States}

\author{Raghav G. Jha}
\email{rjha1@perimeterinstitute.ca}
\affiliation{Perimeter Institute for Theoretical Physics, Waterloo, Ontario N2L 2Y5, Canada}

\author{David Schaich}
\email{david.schaich@liverpool.ac.uk}
\affiliation{Department of Mathematical Sciences, University of Liverpool, Liverpool L69 7ZL, United Kingdom}

\author{Toby Wiseman}
\email{t.wiseman@imperial.ac.uk}
\affiliation{Theoretical Physics Group, Blackett Laboratory, Imperial College, London SW7 2AZ, United Kingdom}

\begin{abstract}
In the large-$N$ and strong-coupling limit, maximally supersymmetric SU($N$) Yang--Mills theory in $(2 + 1)$ dimensions is conjectured to be dual to the decoupling limit of a stack of $N$ D$2$-branes, which may be described by IIA supergravity.
We study this conjecture in the Euclidean setting using nonperturbative lattice gauge theory calculations.
Our supersymmetric lattice construction naturally puts the theory on a skewed Euclidean 3-torus.
Taking one cycle to have anti-periodic fermion boundary conditions, the large-torus limit is described by certain Euclidean black holes.
We compute the bosonic action---the variation of the partition function---and compare our numerical results to the supergravity prediction as the size of the torus is changed, keeping its shape fixed.
Our lattice calculations primarily utilize $N = 8$ with extrapolations to the continuum limit, and our results are consistent with the expected gravity behavior in the appropriate large-torus limit.
\end{abstract}

\maketitle

\section{\label{sec:intro}Introduction}
It has been conjectured~\cite{Maldacena:1997re, Itzhaki:1998dd, Witten:1998zw, Aharony:1999ti} that the large-$N$ limits of maximally supersymmetric Yang--Mills (SYM) theories, obtained from the dimensional reduction of $\cN = 1$ SYM in ten dimensions down to $(p + 1)$ dimensions, are dual to string theories containing D$p$-branes.
In the large-$N$ and strong-coupling limit this relates properties of gauge theories to the dual properties of D$p$-brane solutions in supergravity.
The $p = 3$ case is the AdS/CFT correspondence, which has received much attention, in part due to its additional conformal symmetries.
For direct numerical tests of holographic duality the $p < 3$ cases are more attractive to consider, as they feature more tractable gauge theories~\cite{Schaich:2018mmv}.

For example, the D$0$-brane or $p = 0$ case is a quantum-mechanical description well-known as the Banks--Fischler--Shenker--Susskind (BFSS) model~\cite{deWit:1988wri, Banks:1996vh}.
One of the earliest efforts to understand holographic duality in the quantum-mechanical case directly from non-perturbative gauge theory was described in Refs.~\cite{Kabat:1999hp, Kabat:2000zv, Kabat:2001ve}.
In recent years, good agreement has been obtained for the case of $p = 0$ in the Euclidean setting using numerical Monte Carlo calculations.
These efforts started with Refs.~\cite{Hanada:2007ti, Catterall:2007fp, Anagnostopoulos:2007fw, Catterall:2008yz, Hanada:2008ez, Catterall:2009xn}, and more sophisticated recent lattice analyses give convincing agreement with dual-gravity black hole predictions in the large-$N$ low-temperature limit~\cite{Kadoh:2015mka, Filev:2015hia, Berkowitz:2016tyy, Berkowitz:2016jlq}.
In addition to the BFSS quantum mechanics, a maximally supersymmetric deformation of it known as the Berenstein--Maldacena--Nastase (BMN) model~\cite{Berenstein:2002jq} which may also be dual to black holes at low temperatures~\cite{Costa:2014wya}, is now also starting to be studied on the lattice~\cite{Catterall:2010gf, Asano:2018nol, Schaich:2020ubh}.

This Euclidean lattice approach was extended to the higher-dimensional D$1$-brane case in Refs.~\cite{Catterall:2010fx, Kadoh:2017mcj, Catterall:2017lub, Jha:2017zad}.
To allow numerical lattice calculations, one must compactify the spatial direction.
In the continuum this corresponds to placing the dual theory on a Euclidean torus with all bosonic fields subject to periodic boundary conditions along all directions.
With periodic fermion boundary conditions along all directions, supersymmetry is unbroken and the partition function is independent of the size and shape of this torus.
In order to study more interesting behavior, we take one cycle to be anti-periodic for fermions.

As discussed in Refs.~\cite{Catterall:2017lub,Jha:2017zad}, a conventional thermodynamic interpretation would require the gauge theory to be on a rectangular torus with anti-periodic fermion (thermal) boundary conditions on the Euclidean time cycle.
However, often it is more convenient to work with a skewed torus in the Euclidean setting, in order to use supersymmetric lattice actions which employ non-cubical lattices with enhanced point group symmetries.
While one cannot continue the numerical results to Lorentzian signature due to the skewing, this is not an obstruction to testing supergravity predictions.
One may also consider the dual supergravity in the Euclidean setting with a skewed torus as the asymptotic boundary geometry, in which case in the appropriate large-$N$ 't~Hooft limit it predicts a behavior governed by certain Euclidean black holes (which also have no Lorentzian analog).

The higher-dimensional SYM theories, such as the one considered in this paper, involve more challenging calculations than in the quantum-mechanical case, but offer the advantage of richer structures.
Distinct phases are associated to center symmetry breaking signaled by the eigenvalue distributions of the Wilson lines around the spatial torus cycles, and are described in the dual gravity by the competition between different black hole solutions.
In Refs.~\cite{Catterall:2017lub,Jha:2017zad} these different phases were indeed seen in two-dimensional lattice calculations, and reasonable agreement was observed for the variation of the partition function with torus size for both IIA and IIB supergravity predictions.
(See \refcite{Hiller:2005vf} for an alternate approach to the strong-coupling limit of the $p = 1$ theory in the Lorentzian signature.)

The purpose of this paper is to advance these tests of holographic duality to the next higher dimension --- the case of D$2$-branes.
Again we consider the Euclidean theory compactified on a torus so that it is amenable to lattice calculations.
We take one anti-periodic cycle for fermions.
In the conventional Euclidean thermal setting on a rectangular torus, the system has an even richer phase structure than the case of $p=1$, with sensitivity to the dimensionless temperature and the various aspect ratios of the 3-torus~\cite{Susskind:1997dr,Martinec:1998ja}.
We consider here the skewed torus, as dictated by our supersymmetric lattice discretization.
Keeping the shape of the torus fixed, we vary its size relative to the scale set by the 't~Hooft coupling and study the bosonic action---the variation of the partition function.
We choose the shape of the torus so that we can expect the behavior in the large-$N$ strongly coupled large-torus limit to be governed by the simplest gravitational dual, a homogeneous Euclidean D$2$-brane black hole in IIA supergravity with boundary given by the skewed torus.
We then numerically analyze this large-$N$, large-torus limit, to understand how well the gauge theory matches the predictions of the supergravity solution.

We begin in the next section by discussing $(2 + 1)$-dimensional SYM on a skewed torus and its supergravity dual in the large-$N$ 't~Hooft~limit.
In \secref{sec:lattice} we describe our three-dimensional supersymmetric lattice construction, which produces the numerical results presented and compared with supergravity expectations in \secref{sec:results}.
The data leading to these results are available at~\cite{data}.
We conclude in \secref{sec:conc} by looking ahead to further lattice SYM studies that can build on this work in the future, including prospects for exploring phase transitions by changing the shape of the torus.

\section{\label{sec:continuum}SYM on a skewed torus and the supergravity dual}
We consider three-dimensional maximally supersymmetric Yang--Mills theory, which we take in Euclidean signature to be on a 3-torus denoted hereafter by $\Tbb^3$.
As in the thermal case, we impose anti-periodic fermion boundary conditions only on one cycle corresponding to Euclidean time.
Labelling this coordinate as $\tau$, and the others as $x_i$, we identify $\tau \sim \tau + \be$ (anti-periodic for fermions), while the others form the `spatial' torus cycles after the identifications $(\tau, x_i) \sim (\tau, x_i) + \vec{L}_{1,2}$ (periodic for fermions).
If $\vec{L}_{1,2}$ were orthogonal to each other and $\tau$, the torus would be rectangular and we would have a Lorentzian interpretation with \be being the inverse temperature.
Here we will consider a skewed torus, for which there is no simple Lorentzian interpretation --- the Euclidean torus cannot be analytically continued to a real Lorentzian-signature space-time.
Nonetheless, holographic duality states that this theory can be described by a string theory dual which reduces to supergravity in the large-$N$ 't~Hooft limit.

It is convenient to define dimensionless lengths $r_{\tau} = \be \la$ and $r_{1,2} = |\vec{L}_{1,2}| \la$ in terms of the (dimensionful) 't~Hooft coupling $\la = N g_{\text{YM}}^2$.
Here we are interested in fixing the shape of the torus that the SYM is defined on, while varying its size.
Thus we make the choice $\vec{L}_{1,2} = \be \vec{l}_{1,2}$, with $\vec{l}_{1,2}$ being vectors that we take to be fixed with unit length, $|\vec{l}_{1,2}| = 1$, so that each torus cycle has equal proper length $\be$.
The partition function of the Euclidean theory is then just a function of the one dimensionless parameter $r_{\tau}$, and it is convenient to think in terms of $t = 1 / r_{\tau}$, which we may view as a dimensionless `generalized' temperature.
At large $N$ in the 't~Hooft limit we regard $t \sim O(1)$.
In this limit, a large numerical value $t = 1/ r_{\tau} \gg 1$ corresponds to the torus being small in units of the 't~Hooft coupling, and the theory reduces to a 0-dimensional effective theory of the zero modes on the torus.
This small-torus effective theory corresponds to the bosonic Yang--Mills matrix integral formed from the bosonic truncation of the $p = 0$ SYM theory, which we note is not a weakly coupled description~\cite{Aharony:2004ig, Aharony:2005ew}.

Conversely, a small numerical value $t = 1/ r_\tau \ll 1$ corresponds to the torus being large in units of the 't~Hooft coupling.
The behavior in this regime is given by the decoupling limit of D$2$-branes~\cite{Itzhaki:1998dd}, which may be described in supergravity by the ten-dimensional Euclidean string frame metric and dilaton,
\begin{widetext}
\begin{equation}
  \begin{split}
    ds^2_{\text{IIA, String}} & = \al' \left(\frac{U^{5/2}}{\sqrt{6\pi^2 \la}}\left[\left(1 - \frac{U_0^5}{U^5}\right) d\tau^2 + dx_i^2\right] + \frac{\sqrt{6\pi^2 \la}}{U^{5/2}} \left[dU^2 \left(1 - \frac{U_0^5}{U^5} \right)^{-1} + U^2 d\Omega^2_{(6)}\right]\right) \\
    e^\phi & =  \frac{\la}{N} \sqrt{\frac{6\pi^2 \la}{U^{5/2}}}.
  \end{split}
\end{equation}
\end{widetext}
There is also a 3-form potential carrying the $N$ units of D$2$-charge, with $\tau$ and $x_i$ forming the `world-volume' directions that constitute the asymptotic toroidal boundary which we may think of the gauge theory living on.
Here $U$ is the radial direction, normalized as an energy scale, and $U_0$ represents the radial position of a Euclidean `horizon' where the Euclidean time circle direction, $\tau$, shrinks to zero size.
The smoothness of the geometry relates this to the inverse temperature $\be$, as $U_0^{3/2} = \frac{4\pi^2}{5} \frac{\sqrt{6\la}}{\be}$.
We require large $N$ to suppress string quantum corrections to the supergravity approximation, while the large torus size, $t \ll 1$, is required to suppress the $\al'$ corrections to the classical supergravity geometry near the horizon.
Both these conditions are satisfied if we take $1 \ll r_{\tau} \ll N^{\frac{6}{5}}$ at large $N$, which is the regime we focus on in this work.\footnote{For still-larger tori it is believed the theory flows to a super-conformal IR fixed point given by the Aharony--Bergman--Jafferis--Maldacena (ABJM) model~\cite{Aharony:2008ug} with a dual M$2$-brane description.}

On a large torus with $t \ll 1$, stringy winding modes along the $x_i$ cycles may become relevant, associated to a T-dual Gregory--Laflamme instability~\cite{Gregory:1993vy, Susskind:1997dr, Barbon:1998cr, Li:1998jy, Martinec:1998ja, Aharony:2004ig, Aharony:2005ew}, in the case that $r_{1,2}^{3/2} \lesssim r_{\tau}$.
However, since we are fixing the shape of the torus to have $r_{1,2} = r_{\tau}$, we do not expect such phenomena to occur in a regime where the dual supergravity describes the system.
Since the dual D$2$-brane solution has non-contractible spatial cycles on the torus, we expect the angular distribution of eigenvalues of a Wilson line about such a cycle to be homogeneous at large $N$~\cite{Maldacena:1998im, Witten:1998zw, Aharony:2004ig}.
On the other hand, for a small torus where the theory reduces to a bosonic matrix integral, we expect a highly localized distribution of eigenvalue phases for Wilson lines about any torus cycle.
Hence one expects a large-$N$ transition as the torus size is varied, associated to center symmetry breaking of the spatial Wilson lines.

If the $x_i$ directions were not compact, so that $T = 1 / \be$ is a temperature, then noting that the solution is translation invariant in the $\tau$ and $x_i$ directions, one may compute the free energy density $f$ from the dual-gravity solution,
\begin{align}
  \frac{f}{N^2 \la^3} & = -\left(\frac{2^{13} 3^5 \pi^8}{5^{13}}\right)^{1 / 3} t^{10 / 3} \approx -2.49189 ~ t^{10 / 3} \, .
\end{align}
Compactifying on a torus doesn't change this density, and for a rectangular torus it yields a partition function $\log Z = - f V(\Tbb^3)$, where $V(\Tbb^3)$ denotes the volume of the 3-torus.
Due to the translation invariance of the solution, the skewed-torus partition function is given by these same expressions, although there is no thermal interpretation~\cite{Aharony:2005ew}.

The SYM action is composed of bosonic and fermionic parts having the schematic form
\begin{align}
  S_{\text{SYM}} & = S_{\text{Bos}} + S_{\text{Ferm}} \\
  S_{\text{Bos}} & = \frac{N}{4\la} \int_{\Tbb^3} d\tau d^2x \, \Tr{F^2 + 2\left(D \Phi_I\right)^2 - \left[\Phi_I, \Phi_J \right]^2} \cr
  S_{\text{Ferm}} & = \frac{N}{\la} \int_{\Tbb^3} d\tau d^2x \, \Tr{\psi^T \left(\slashed{D} - \left[\Ga_I \Phi_I, \cdot \right] \right) \psi}. \nonumber
\end{align}
Rescaling the gauge field $A$, scalars $\Phi_I$, fermions $\psi$, and the coordinates $(\tau, x_i)$ by the torus size so they are all dimensionless,
\begin{align*}
  (A, \Phi_I) & = (A', \Phi_I') / \be &
         \psi & = \sqrt{\la} \psi' / \be \\
  (\tau, x_i) & = \be (\tau', x_i'), & &
\end{align*}
the action may be written as
\begin{equation}
  S_{\text{SYM}} = \frac{1}{\be \la} S'_{\text{Bos}} + S'_{\text{Ferm}},
\end{equation}
where $S'_{\text{Bos}} = S_{\text{Bos}} \be \la$ and $S'_{\text{Ferm}} = S_{\text{Ferm}}$ involve only the dimensionless bosonic fields and fermion fields respectively, and have no explicit \be or \la dependence.
Thus we may explicitly differentiate the partition function with respect to \be to obtain
\begin{equation}
  \be \frac{\partial}{\partial \be} \log Z = \vev{S_{\text{Bos}}}.
\end{equation}
While the partition function itself cannot be computed through the lattice methods we use, the expectation value of the bosonic action is very convenient to obtain (as reviewed in the Appendix).
We find the prediction from supergravity that at large $N$,
\begin{align}
  \frac{\vev{S_{\text{Bos}}}}{N^2} & = -\left(\frac{2^{13} 3^2 \pi^8 t}{5^{13}}\right)^{1 / 3} \left(\frac{V(\Tbb^3)}{\be^3}\right)
\end{align}
when $t$ is sufficiently small.
In the small-volume limit $t \gg 1$ we may use the effective dimensional reduction to compute
\begin{equation}
  \frac{\vev{S_{\text{Bos}}}}{N^2} = - 2
\end{equation}
at large $N$~\cite{Kawahara:2007ib,Catterall:2017lub}.

We will see in the next section that the most natural torus geometry for us to consider is formed by periodically identifying $\Rbb^3$ in the three basis directions of an $A_3^*$ lattice.
As discussed above, we do so taking the cycle in each direction to have the same length $\be$.
Explicitly in our coordinates $x_{\mu} = (\tau, x_i)$ we may achieve this by taking
\begin{align}
  \vec{l}_1 & = \frac{1}{3} \left(\begin{array}{c}-1 \\
                                                   2\sqrt{2} \\
                                                   0\end{array}\right) &
  \vec{l}_2 & = \frac{1}{3} \left(\begin{array}{c}-1 \\
                                                  -\sqrt{2} \\
                                                  \sqrt{6}\end{array}\right),
\end{align}
which gives a volume $V(\Tbb^3) = 4\be^3 / (3\sqrt{3})$.

Defining the bosonic action density $s_{\text{Bos}} = \vev{S_{\text{Bos}}} / V(\Tbb^{3})$, for our torus geometry we see the holographic large-volume behavior and small-volume limit imply
\begin{equation}
  \label{eq:limits}
  \frac{s_{\text{Bos}}}{N^2 \la^3} = \left\{\begin{array}{ll}-0.831\dots t^{10 / 3} & \mbox{for } t \ll 1 \\
                                                             -2.598\dots t^3        & \mbox{for } t \gg 1\end{array}\right. .
\end{equation}
It is worth noting that for SYM on an analogous torus in $(p + 1)$-dimensions we would have parametric dependence $s_{\text{Bos}} \propto t^{(14-2p)/(5-p)}$ for $t \ll 1$ from the gravity dual, and the $t \gg 1$ limit would go as $s_{\text{Bos}} \propto t^{p + 1}$.
In the $p = 3$ conformal case these powers coincide, and we see the powers in the case of $p=2$ we consider here are rather close.
This makes the task of distinguishing the two behaviors more challenging than for the $p = 0$ and $1$ cases considered previously~\cite{Hanada:2008ez, Catterall:2008yz, Berkowitz:2016tyy, Berkowitz:2016jlq, Kadoh:2017mcj, Catterall:2017lub, Jha:2017zad}, where there is greater contrast between the large- vs.\ small-volume parametric dependence on $t$.

\section{\label{sec:lattice}Three-dimensional supersymmetric lattice construction}
In recent years, it has become possible to formulate certain supersymmetric lattice gauge theories using the idea of topological twisting, in which the supercharges are grouped into $p$-forms and the $0$-form supercharges can be preserved in discrete space-time.
While this construction is not needed for $(0 + 1)$-dimensional SYM quantum mechanics (where one can show perturbatively that no relevant supersymmetry-breaking counter-terms are possible~\cite{Giedt:2004vb, Catterall:2007fp}), in higher dimensions it is a key ingredient to minimize issues of fine-tuning~\cite{Catterall:2009it, Schaich:2018mmv}.

The three-dimensional maximally supersymmetric Yang--Mills theory considered here can be obtained by classical dimensional reduction of four-dimensional $\cN = 4$ SYM.
The $\cN = 4$ SYM lattice construction~\cite{Kaplan:2005ta, Catterall:2007kn, Damgaard:2008pa, Catterall:2012yq, Catterall:2013roa, Catterall:2014vka, Catterall:2014mha, Schaich:2014pda, Catterall:2015ira} discretizes a maximal twist of the continuum theory known as the Marcus or geometric-Langlands twist~\cite{Marcus:1995mq, Kapustin:2006pk}.
The resulting lattice theory features many symmetries: in addition to U($N$) lattice gauge invariance and a single scalar supersymmetry, it is also invariant under a large $S_5$ point group symmetry arising from the underlying $A_4^*$ lattice.
Using these symmetries, it is possible to show in perturbation theory that radiative corrections generate only a small number of log divergences in the lattice theory~\cite{Catterall:2013roa}.
On reduction to three dimensions these divergences disappear and no fine-tuning is expected to be needed to take the continuum limit~\cite{Kaplan:2005ta}.
The resulting three-dimensional lattice theory naturally lives on an $A_3^*$ (body-centered cubic) lattice, whose four basis vectors correspond to vectors drawn out from the center of an equilateral tetrahedron to its vertices.

As we did in Refs.~\cite{Catterall:2017lub,Jha:2017zad}, here we use the full four-dimensional lattice construction provided by the publicly available parallel software described in Refs.~\cite{Schaich:2014pda, parallel_imp},\footnote{{\tt\href{https://github.com/daschaich/susy}{github.com/daschaich/susy}}} setting $N_z = 1$ to reduce to the $A_3^*$ lattice.
The remaining lattice directions are taken to have equal numbers of lattice sites, $N_x = N_y = N_{\tau}$, with anti-periodic fermion boundary conditions only on the $N_{\tau}$ cycle.
In the continuum limit this generates the skewed torus geometry described in \secref{sec:continuum} (re-labelling $\{x_1, x_2\}$ as $\{x, y\}$).

We relegate the full details of the lattice action $S_{\text{lattice}}$ to the Appendix, and here discuss only the two soft-supersymmetry-breaking deformation that need to be included in order to enable our three-dimensional numerical computations.
The first of these is a scalar potential term, which regulates the divergences associated with integration over a non-compact moduli space in the partition function.
We have used various scalar potentials in our previous investigations, and here employ the single-trace version also used in Refs.~\cite{Catterall:2017lub, Jha:2017zad}:
\begin{equation}
  \label{eq:single_trace}
  S_{\text{soft}} = \frac{N}{4\lalat} \mu^2 \sum_{\vn, a} \Tr{\bigg(\cUbar_a(\vn) \cU_a(\vn) - \Ibb_N\bigg)^2},
\end{equation}
with $\mu^2$ a tunable coefficient and the dimensionless \lalat defined in the appendix. 
We need to extrapolate $\mu^2 \to 0$ in order to recover the continuum SYM theory of interest, in addition to extrapolating to the continuum limit of vanishing lattice spacing that corresponds to $\lalat \to 0$ in fewer than four dimensions.
We guarantee that $\mu^2 \to 0$ in the $\lalat \to 0$ continuum limit by setting $\mu = \zeta \lalat$.
This also allows us to extrapolate $\mu^2 \to 0$ with \lalat fixed by considering the $\zeta^2 \to 0$ limit, which we will do in \secref{sec:results}.

Next, for the dimensionally reduced lattice theory to correctly reproduce the continuum physics, we need to ensure that the trace of the each gauge link $\cU_z(\vn)$ in the reduced $z$-direction is close to $N$, so that the effective scalar field obtained by dimensional reduction is small in lattice units.
In other words, this means that the center symmetry should be completely broken in the reduced direction for proper dimensional reduction.
We ensure this by adding a second soft-supersymmetry-breaking deformation to the lattice action:
\begin{equation}
  \label{eq:center}
  S_{\text{center}} = \frac{N}{4\lalat} \ka^2 \sum_{\vn} \mbox{Tr}\bigg[\bigg(\cU_z(\vn) - \Ibb_N\bigg)^{\dag}\bigg(\cU_z(\vn) - \Ibb_N\bigg)\bigg],
\end{equation}
with $\ka^2$ another tunable coefficient that we must also take to zero in \secref{sec:results}.
This term is gauge invariant since $N_z = 1$.
It explicitly breaks the center symmetry in the single reduced direction by forcing the trace of the link in this direction to be close to $N$.

With this lattice action $S_{\text{lattice}}$ for three-dimensional SYM, we stochastically sample field configurations using the rational hybrid Monte Carlo (RHMC) algorithm~\cite{Clark:2006fx} implemented in the software mentioned above~\cite{Schaich:2014pda, parallel_imp}.
The RHMC algorithm treats $e^{-S_{\text{lattice}}}$ as a Boltzmann weight, requiring that we consider a lattice action that is real and non-negative.
However, gaussian integration over the fermion fields of three-dimensional SYM produces a pfaffian that is potentially complex,
\begin{equation}
  \int \left[d\Psi\right] e^{-\Psi^T \cD \Psi} \propto \pf \cD = |\pf \cD| e^{i\phi}.
\end{equation}
Here \cD is the fermion operator and $S_{\text{lattice}} = S_{\text{Bos}} + \Psi^T \cD \Psi$, with $S_{\text{Bos}}$ the bosonic part of the lattice action.

As in our previous work~\cite{Catterall:2014vka, Catterall:2014vga, Catterall:2015ira, Schaich:2015daa, Catterall:2017lub, Jha:2017zad, Schaich:2018mmv}, we `quench' the phase $e^{i\phi} \to 1$ to obtain a positive lattice action for use in the RHMC algorithm.
Reweighting
\begin{equation}
  \vev{\cO} = \frac{\vev{\cO e^{i\phi}}_{\text{pq}}}{\vev{e^{i\phi}}_{\text{pq}}}
\end{equation}
is then required to recover expectation values from these phase-quenched (`$_{\text{pq}}$') calculations, where
\begin{align}
  \vev{\cO}_{\text{pq}} & = \frac{\int[d\cU] \ \cO e^{-S_{\text{Bos}}}\ |\pf \cD|}{\int[d\cU] \ e^{-S_{\text{Bos}}} \ |\pf \cD|} \\
  \vev{\cO}             & = \frac{\int[d\cU] \ \cO e^{-S_{\text{Bos}}}\ \pf \cD}{\int[d\cU] \ e^{-S_{\text{Bos}}} \ \pf \cD}.
\end{align}
This procedure breaks down, producing a sign problem, when $\vev{e^{i\phi}}_{\text{pq}}$ is consistent with zero.
Fortunately, in this investigation we focus on regimes where $\vev{e^{i\phi}}_{\text{pq}} \approx 1$ and $\vev{\cO} \approx \vev{\cO}_{\text{pq}}$.
This follows from the fact that the $r_{\tau}$ and $N_{\tau}$ we analyze correspond to $0.14 < \lalat < 1.34$, safely in the range of couplings where we observe $\vev{e^{i\phi}}_{\text{pq}} \approx 1$ in the full four-dimensional theory~\cite{Catterall:2014vga, Schaich:2015daa, Schaich:2018mmv}.\footnote{These calculations used a double-trace scalar potential in place of \eq{eq:single_trace}, which should not noticeably affect pfaffian phase fluctuations.}
In addition, we gain further benefit from the dimensional reduction, since the lower-dimensional continuum limit corresponds to $\lalat \to 0$.
Partly for this reason, previous lattice studies of $\cN = (2, 2)$ and $\cN = (8, 8)$ SYM theories in two dimensions found $\vev{e^{i\phi}}_{\text{pq}} \to 1$ rapidly upon approaching the continuum limit, with negligible pfaffian phase fluctuations even at non-zero lattice spacing~\cite{Hanada:2010qg, Catterall:2011aa, Mehta:2011ud, Galvez:2012sv, Catterall:2017xox}.
Similarly small pfaffian phase fluctuations were also seen in the $p = 0$ case~\cite{Catterall:2009xn, Filev:2015hia}.

\section{\label{sec:results}Numerical results and comparison with supergravity}
\begin{figure*}[tbp]
  \includegraphics[width=0.48\linewidth]{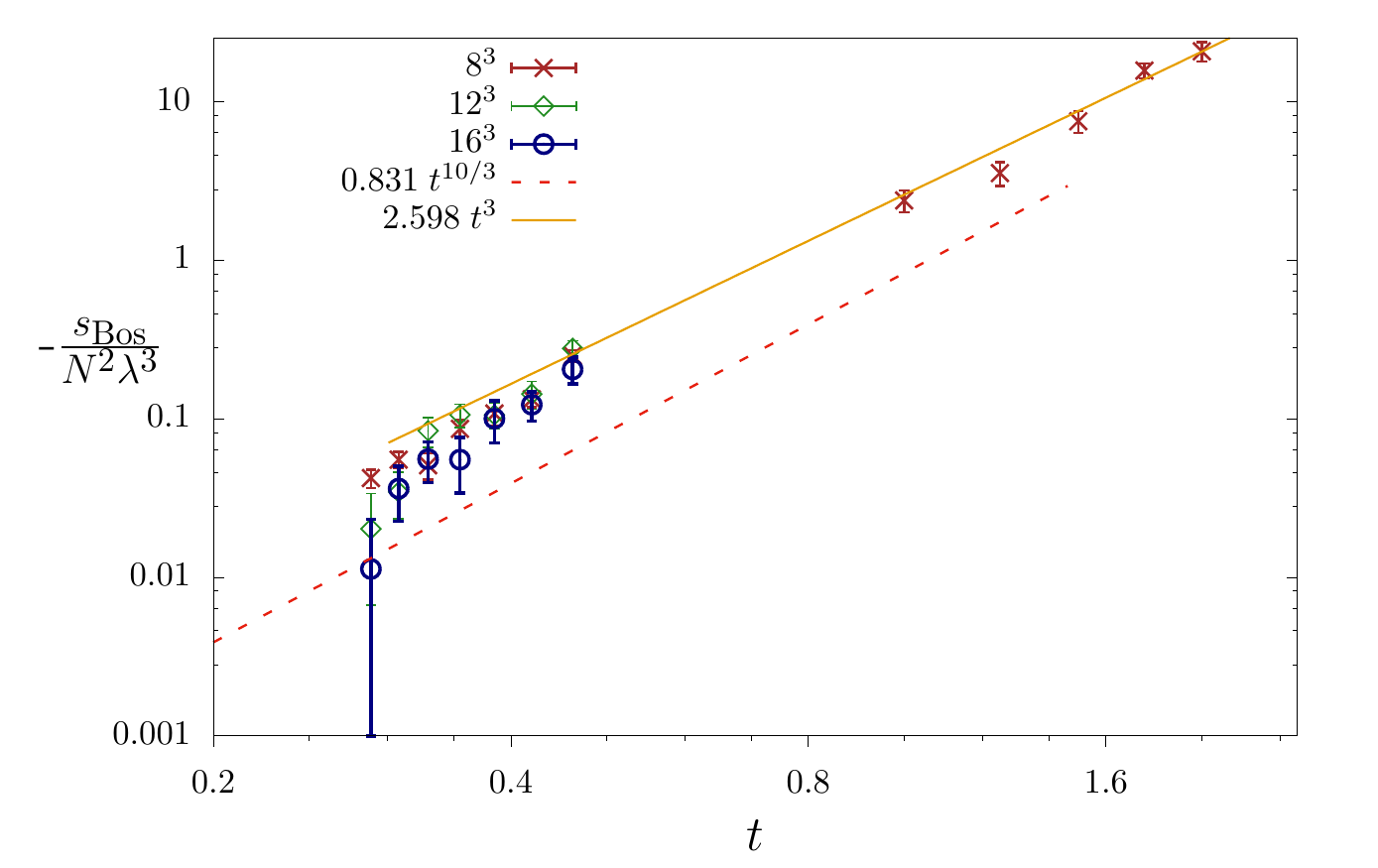}\hfill \includegraphics[width=0.48\linewidth]{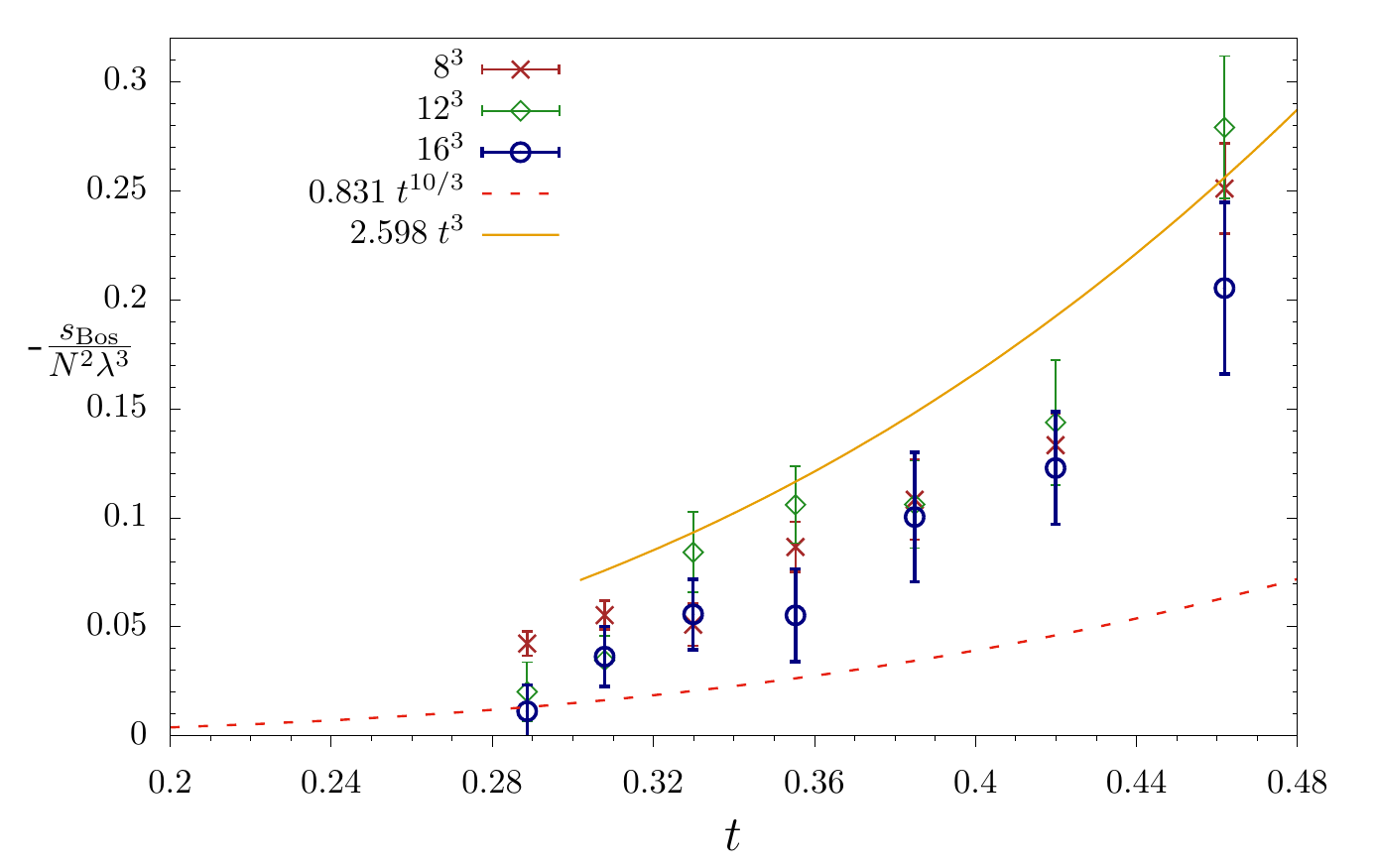}
  \caption{\label{fig:main}The $\zeta^2 \to 0$ extrapolated bosonic action density for $N = 8$ with lattice sizes $8^3$, $12^3$, and $16^3$, compared with the large-volume (dashed) and small-volume (solid) expectations from \protect\eq{eq:limits}.  \textbf{Left:} The full range of dimensionless temperatures $t$ on log--log axes.  \textbf{Right:} Focusing on $ 0.2 < t < 0.48$  with linear axes to clarify the absolute size of uncertainties.}
\end{figure*}

We now present our lattice results for the bosonic action density in the two different regimes described in \secref{sec:continuum}.
Recall that the small-volume regime has dimensionless `generalized' temperature $t \gg 1$, while $t \ll 1$ for the more interesting large-volume regime related to the dual supergravity by holography.
We have concentrated resources to analyze these two regimes, with a focus on $0.25 < t < 0.5$.
Our key result is \fig{fig:main} where we display the bosonic action density vs.\ $t$ for $N = 8$ and the $L^3$ lattice sizes we consider, with $N_x = N_y = N_{\tau} = L = 8$, $12$, and $16$.

After briefly discussing $t \geq 1$ results in the small-volume regime, which we use to check our lattice calculations, we focus on the more challenging large-volume case with $0.25 < t < 0.5$.
This range of $t$ is chosen to satisfy the conditions $1 \ll r_{\tau} \ll N^{\frac{6}{5}}$ discussed in \secref{sec:continuum}, which for $N = 8$ correspond to $0.08 \ll t \ll 1$.
While it would be straightforward to run numerical calculations with smaller $t \lsim 0.25$, for our current $N = 8$ these may exit the regime in which IIA supergravity is a reliable description of the holographically dual gravitational system.
Moving to larger $N > 8$ is also possible, but would demand much more substantial computational resources due to computational costs increasing more rapidly than $N^3$~\cite{parallel_imp}.
The results presented here required $\sim$$5$ million core-hours provided by multiple computing facilities, with costs dominated by the largest $L = 16$ we consider. 
Ref.~\cite{data} provides a comprehensive release of our data, including full accounting of statistics, auto-correlation times, extremal eigenvalues of the fermion operator (which must remain within the spectral range where the rational approximation used in the RHMC algorithm is reliable), and other observables computed in addition to the bosonic action density.

\subsection{\label{sec:small}Small-volume regime, $t \gg 1$}
To check that our lattice calculations reproduce the expected small-volume behavior of three-dimensional SYM, we analyze several large values of $t \geq 1$.
Motivated by the right panel of \fig{fig:main}, which shows no significant dependence on $L \geq 8$ for $t \gsim 0.3$, we carry out these calculations for a single $L^3$ lattice size with $L = 8$.
For these large $t$ we are also able to set $\ka^2 = 0$ in \eq{eq:center} without encountering numerical instabilities (i.e., the center symmetry in the reduced direction breaks dynamically), leaving \eq{eq:single_trace} the only soft-supersymmetry-breaking deformation in the lattice action.
As discussed in \secref{sec:lattice}, we remove this deformation by extrapolating $\zeta^2 \to 0$, here considering $\zeta^2 = 0.04$, $0.06$ and $0.09$ for each value of $t$.
These linear extrapolations produce the $t \geq 1$ results in the left panel of \fig{fig:main}, which are in good agreement with the solid line showing the expected small-volume limit from \eq{eq:limits}.

\begin{figure}[tbp]
  \includegraphics[width=\linewidth]{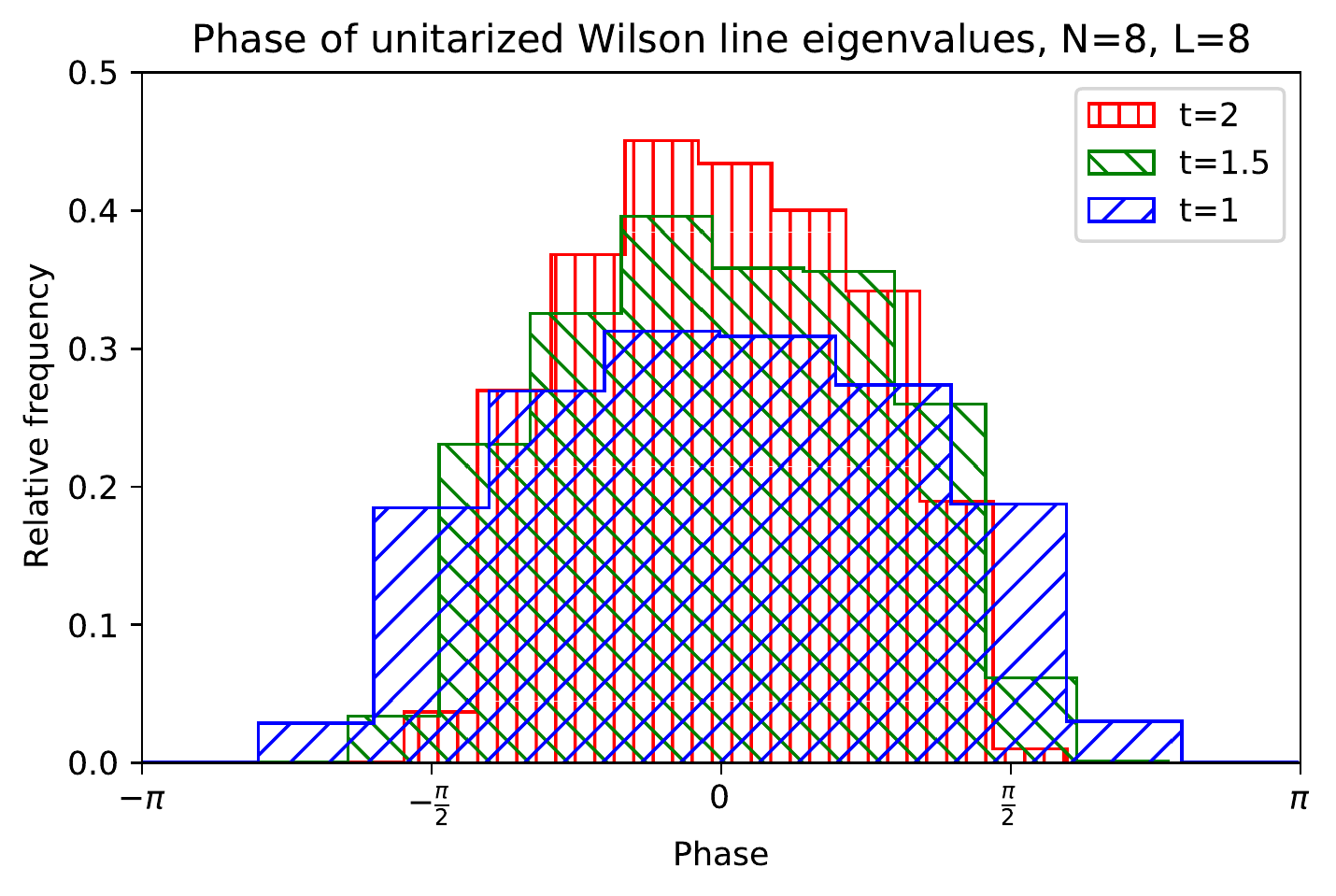}
  \caption{\label{fig:hist_highT}Distributions of $N = 8$ Wilson line eigenvalue phases over the angular range $[-\pi, \pi)$, in the small-volume regime with dimensionless temperatures $t \geq 1$.  The distributions become more localized with increasing $t$, as expected.}
\end{figure}

In \fig{fig:hist_highT} we show distributions of the phases of the Wilson line (spatial holonomy) eigenvalues for three $t \geq 1$ lattice ensembles with $\zeta^2 = 0.09$.
As reviewed in the Appendix, our lattice construction naturally provides complexified Wilson lines that include contributions from both the gauge and scalar fields.
In this work, we remove the scalar-field contributions by considering instead unitarized Wilson lines.
The resulting distributions shown in \fig{fig:hist_highT} are clearly localized, and the width of the support decreases as $t$ increases.
This lattice result is consistent with the expectation that the angular eigenvalue distribution is highly localized for $t \to \infty$, providing another non-trivial check that our lattice calculations correctly reproduce the three-dimensional SYM theory.

\subsection{Large-volume regime, $t \ll 1$}
Turning now to the more interesting large-volume regime where we can compare our results with dual supergravity predictions, we analyze $0.25 < t < 0.5$ in order to satisfy the conditions $N^{-\frac{6}{5}} \ll t \ll 1$ discussed above, with $N^{-\frac{6}{5}} \approx 0.08$ for the $N = 8$ we consider.
In this regime, we need to include both soft-supersymmetry-breaking deformations Eqs.~\ref{eq:single_trace} and \ref{eq:center} in the lattice action.
To simplify our analysis we set $\ka^2 = \mu^2$, so that each $\zeta^2 \to 0$ extrapolation (here considering $\zeta^2 = 0.01$, $0.04$, and $0.09$) simultaneously removes both deformations.
Control over these extrapolations is essential to precisely determine the SYM bosonic action density to be compared with the supergravity prediction.

\begin{figure}[tbp]
  \includegraphics[width=\linewidth]{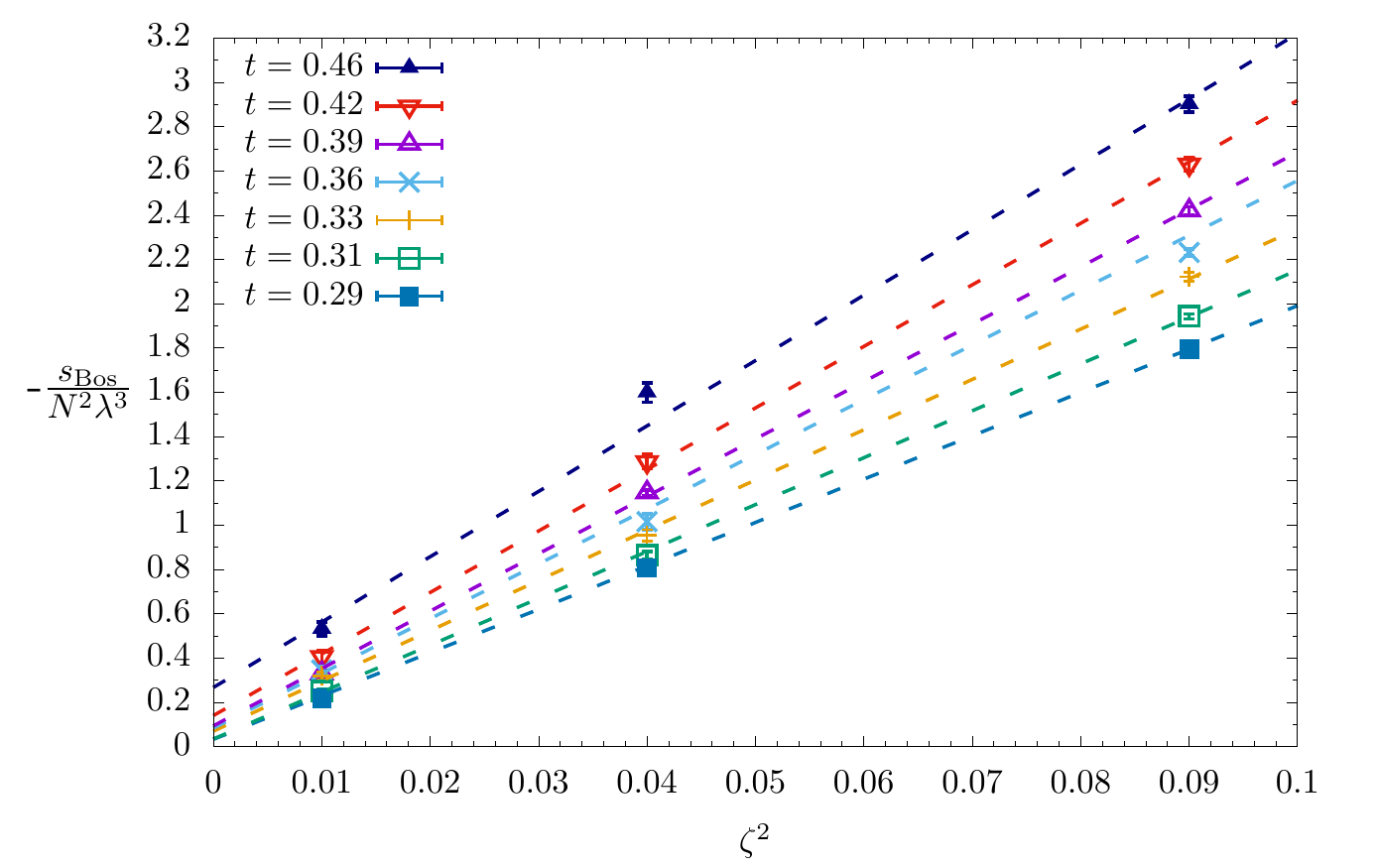}
  \caption{\label{fig:zeta_extrap}Representative linear $\zeta^2 \to 0$ extrapolations of the bosonic action density for different temperatures on $12^3$ lattices with $N = 8$.}
\end{figure}

Representative linear $\zeta^2 \to 0$ extrapolations of our bosonic action density data are shown in \fig{fig:zeta_extrap} for all our $12^3$ lattice ensembles with $N = 8$.
The $\zeta^2 \to 0$ limits in this figure correspond exactly to the $12^3$ points shown in both panels of \fig{fig:main}.
Clearly the $\zeta^2 \to 0$ extrapolated results in \fig{fig:main} have significantly larger relative uncertainties than the input data at non-zero $\zeta$ in \fig{fig:zeta_extrap}.
This is a consequence of the steep extrapolations to the much smaller SYM bosonic action densities that remain after removing the deformations in our lattice action.

\begin{figure}[tbp]
  \includegraphics[width=\linewidth]{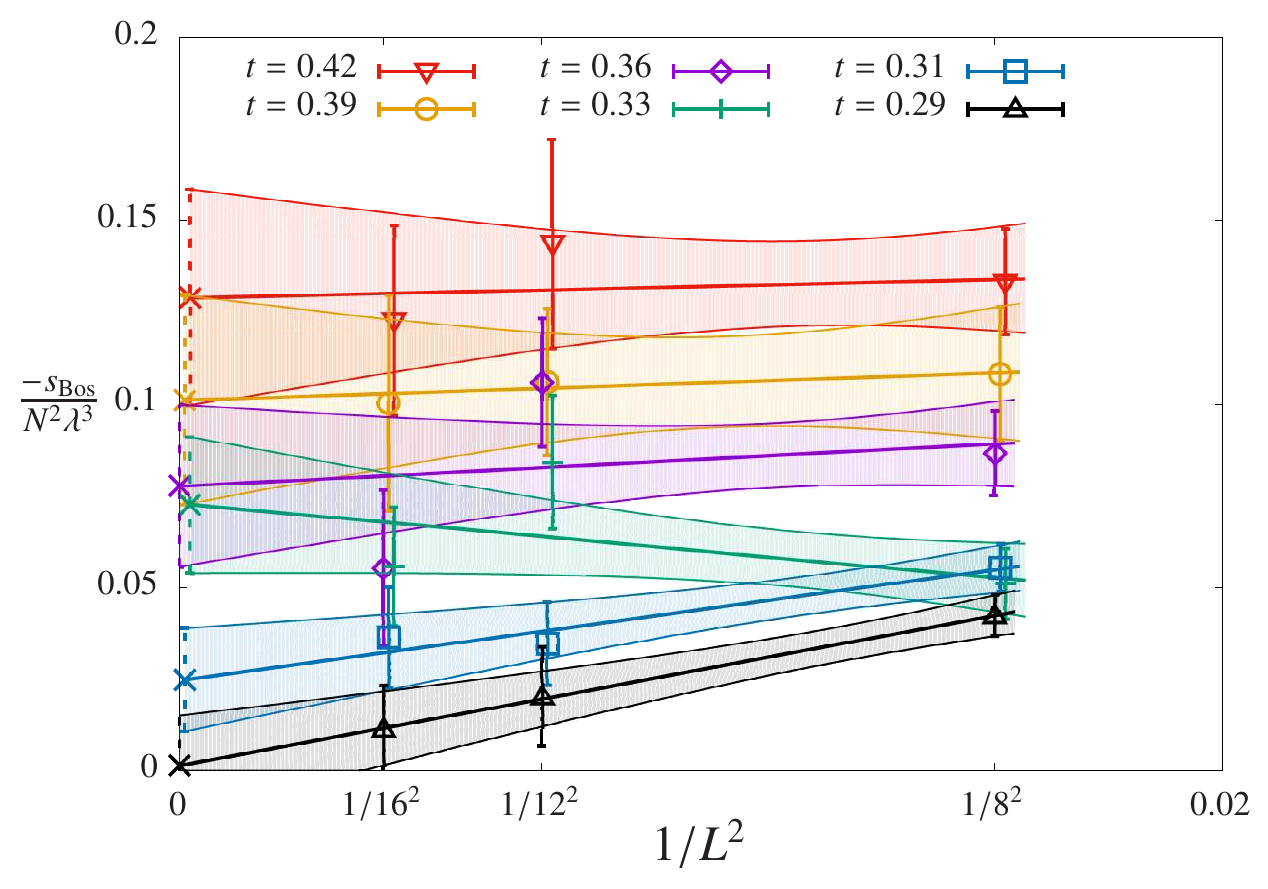}
  \caption{\label{fig:cont_extrap}Continuum extrapolations of the $\zeta^2 \to 0$ extrapolated bosonic action density for various temperatures with $N = 8$, where the limit $L^2 \to \infty$ with fixed $t$ corresponds to $\lalat \to 0$ and vanishing lattice spacing.  Small horizontal offsets are added for clarity.  All extrapolations for $t \geq 0.33$ have slopes consistent with zero, indicating no significant discretization artifacts in the corresponding results.}
\end{figure}

These larger uncertainties are even more evident in \fig{fig:cont_extrap}, where we zoom in on the six smallest $0.29 \lsim t \lsim 0.42$ to investigate the dependence of the $\zeta^2 \to 0$ extrapolated bosonic action densities on the $L^3$ lattice volume with $L = 8$, $12$ and $16$.
Since we fix the dimensionless lengths of the lattice, $r_x = r_y = r_{\tau}$, larger values of $L$ correspond to smaller lattice spacings, allowing us to check discretization artifacts and extrapolate to the continuum limit, $1 / L^2 \to 0$ or equivalently $L^2 \to \infty$.
Most of the linear $1 / L^2 \to 0$ extrapolations shown in \fig{fig:cont_extrap} have slopes consistent with zero, indicating that there are not significant discretization artifacts in the corresponding results, and motivating our choice to include all our $L = 8$, $12$ and $16$ results in \fig{fig:main}.
On the whole, these bosonic action density results are reasonably consistent with the large-$N$ prediction from supergravity in \eq{eq:limits} (the dashed line in \fig{fig:main}), particularly considering the modest $N = 8$ and $t \approx 0.3$ that we have used in this work.

The best agreement with the dual supergravity prediction comes from the two smallest $t \approx 0.31$ and $0.29$, which are also the cases where the $1 / L^2 \to 0$ continuum extrapolations are non-trivial.
From \fig{fig:cont_extrap} we can see that these non-trivial extrapolations are driven by the $L = 8$ results, with the $L = 12$ and $16$ results fully consistent with the respective continuum limits within their (relatively large) uncertainties.
An obvious question in this context is whether these results really fall in the large-volume regime, or may still be governed by small-volume (or intermediate) behavior.
As discussed below \eq{eq:limits}, the expected parametric dependence of the bosonic action density is rather similar in both regimes for this $p = 2$ case, making it more difficult to distinguish a clear change in behavior.

\begin{figure*}[tbp]
  \includegraphics[width=0.48\linewidth]{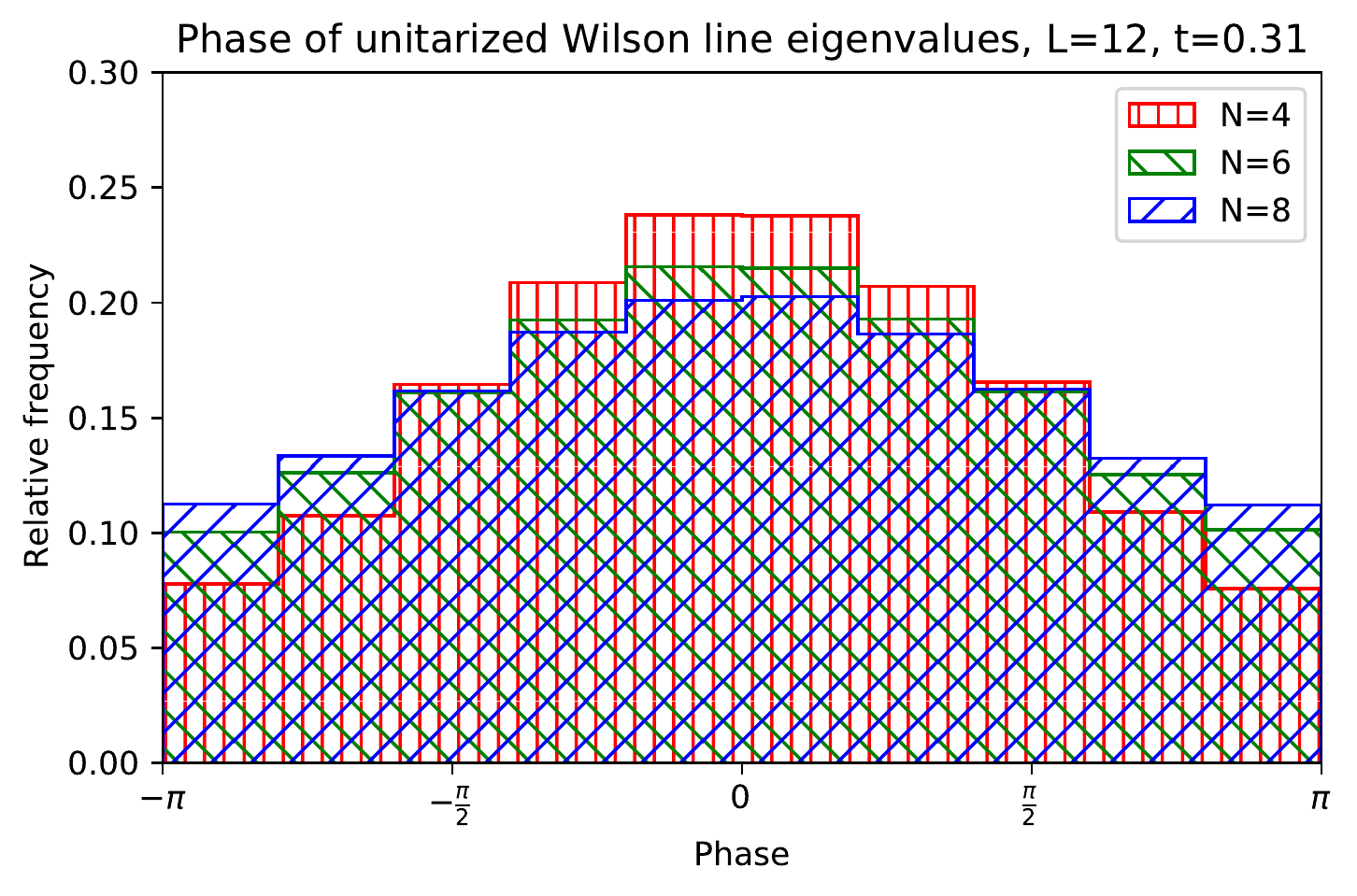}\hfill \includegraphics[width=0.48\linewidth]{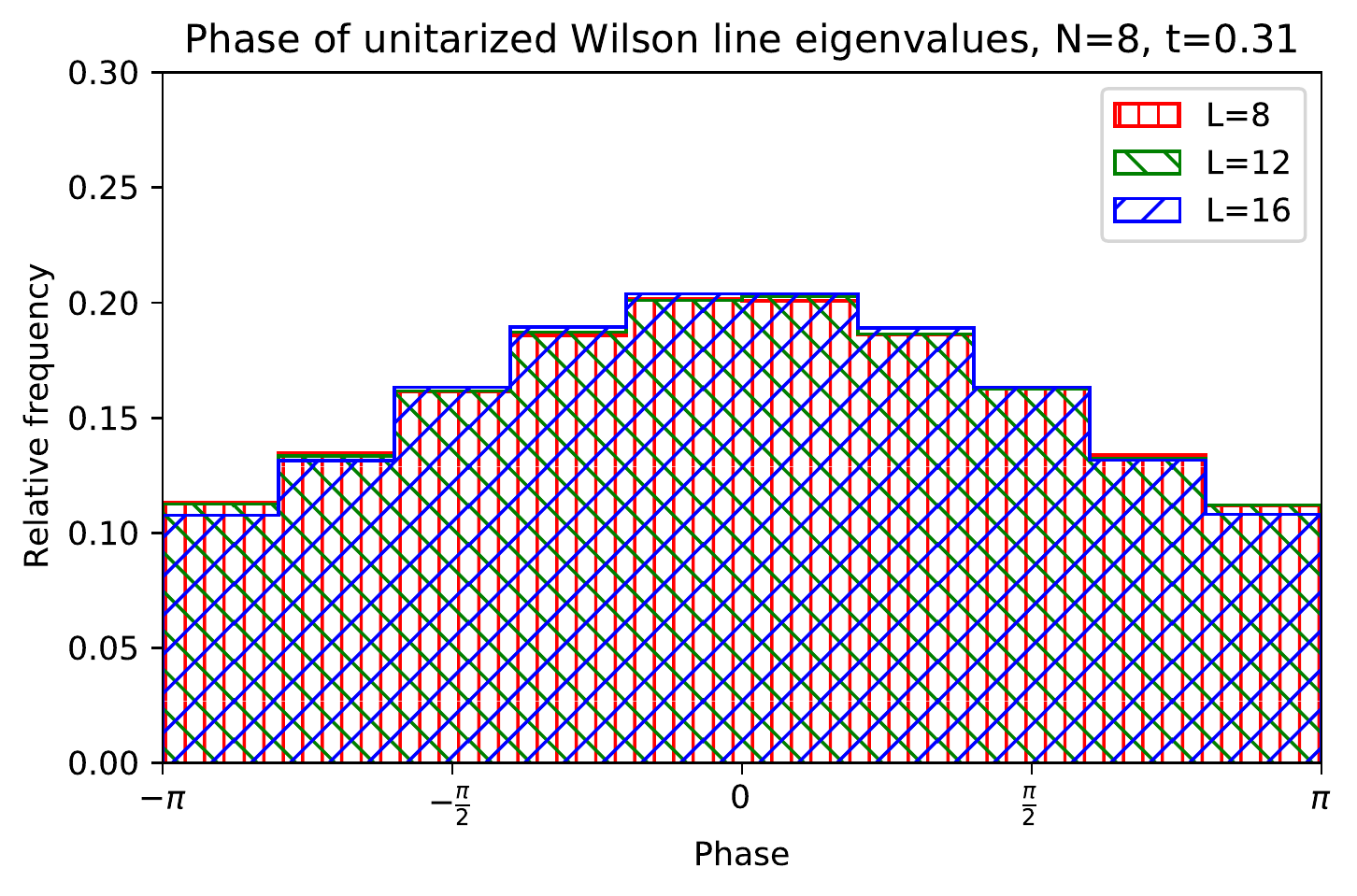}
  \caption{\label{fig:hist_lowT}Distributions of Wilson line eigenvalue phases over the angular range $[-\pi, \pi)$ for a small $t \approx 0.31$.  \textbf{Left:} The $L = 12$ distributions become broader as $N$ increases, consistent with the homogeneous distribution expected for the large-volume regime in the large-$N$ limit.  \textbf{Right:} The $N = 8$ distributions are independent of the lattice size $L^3$.} 
\end{figure*}

Stronger evidence that our small-$t$ results are in the large-volume regime can be obtained by again considering the eigenvalues of the Wilson line about the spatial torus cycles.
In \fig{fig:hist_lowT} we show distributions of the phases of these eigenvalues for lattice ensembles with $t \approx 0.31$ and $\zeta^2 = 0.09$, which follow broad distributions in clear contrast to the small-volume case shown in \fig{fig:hist_highT}.
Recall that the D$2$ supergravity solution predicts a homogeneous distribution of these phases at large $N$.
To check the dependence on $N$, we have generated one $12^3$ ensemble with $N = 4$ and another with $N = 6$.
In the left panel of \fig{fig:hist_lowT} we compare the resulting $N = 4$, $6$ and $8$ Wilson line eigenvalue phase distributions and confirm that they become broader as $N$ increases, consistent with the expected large-$N$ homogeneous distribution.
In the right panel we check that there is no visible $L$ dependence in our $N = 8$ results for this same $t \approx 0.31$ and $\zeta^2 = 0.09$.
Thus we confirm that our small-$t$ results do indeed appear to be in the large-volume regime and consistent with the dual supergravity predictions.
Presumably there is a large-$N$ phase transition separating the small- and large-volume regimes, although such a transition is difficult to see in our $N = 8$ data on the lattice sizes we consider here.

\section{\label{sec:conc}Conclusions and next steps}
We have presented the first numerical lattice gauge theory studies of three-dimensional maximally supersymmetric Yang--Mills theory, advancing our program of non-perturbatively testing holography.
Such tests provide direct first-principles checks of holographic duality at finite temperatures and in non-conformal settings, where tools such as integrability and supersymmetric localization are not available.

Already at modest $N = 8$ our results indicate that the large-$N$ predictions of the dual-gravity black holes can emerge for large tori.
We have seen that the bosonic action density interpolates rather smoothly between the small-volume regime and the large-volume supergravity regime, similar to results for lower-dimensional cases~\cite{Hanada:2008ez, Catterall:2008yz, Berkowitz:2016tyy, Berkowitz:2016jlq, Kadoh:2017mcj, Catterall:2017lub, Jha:2017zad}.
We are able to see qualitative agreement with the supergravity prediction derived from the dual black hole action density, and continuum extrapolations indicate no significant discretization artifacts for $t \geq 0.33$.
We also see that the Wilson lines about the spatial directions of the torus are consistent with a transition from a localized angular eigenvalue distribution at small volumes to the expected homogeneous distribution at large volumes, presumably with a large-$N$ phase transition at an intermediate torus size.

In the future, we plan to look at the Maldacena--Wilson loop and compare it to the results obtained from the dual-gravity computations.
In addition, similar to our previous study~\cite{Catterall:2017lub, Jha:2017zad}, we can also change the aspect ratios of the torus cycle sizes to study phase transitions from the homogeneous D$2$-phase we consider here to D$1$-phases or even localized D$0$-phases.
It will also be interesting to understand the nature of the large-$N$ phase transition at intermediate volumes, although this has proved difficult to study even in simpler settings~\cite{Bergner:2019rca}.

Though our results approach the supergravity predictions in the appropriate regime, even larger $N$ would help to better satisfy the conditions on the validity of the classical supergravity description.
Numerical calculations at larger $N$ are certainly possible, but would require much more substantial computational resources due to computational costs increasing more rapidly than $N^3$~\cite{parallel_imp}.
Our current results in this paper nevertheless show the approach to this regime in detail and are certainly consistent with the supergravity results.

\section*{Acknowledgements}
This work was supported by the US Department of Energy (DOE), Office of Science, Office of High Energy Physics, under Award Numbers {DE-SC0009998} (SC) and {DE-SC0013496} (JG).
RGJ's research is supported by postdoctoral fellowship at the Perimeter Institute for Theoretical Physics.
Research at Perimeter Institute is supported in part by the Government of Canada through the Department of Innovation, Science and Economic Development Canada and by the Province of Ontario through the Ministry of Colleges and Universities.
DS was supported by UK Research and Innovation Future Leader Fellowship {MR/S015418/1}.
Numerical calculations were carried out at the University of Liverpool, on DOE-funded USQCD facilities at Fermilab, and at the San Diego Computing Center through XSEDE supported by National Science Foundation grant number {ACI-1548562}. 

\begin{appendix}
\section*{Appendix: Lattice action and computation of the bosonic action}
Our lattice formulation of maximally supersymmetric Yang--Mills theory in $d < 4$ dimensions discretized on the $A_d^*$ lattice is obtained by classical dimensional reduction from the parent four-dimensional theory.
The lattice action for topologically twisted $\cN = 4$ SYM in $d = 4$ dimensions is the sum of the following $\cQ$-exact and $\cQ$-closed terms~\cite{Kaplan:2005ta, Catterall:2007kn, Damgaard:2008pa, Catterall:2012yq, Catterall:2013roa, Catterall:2014vka, Catterall:2014mha, Schaich:2014pda, Catterall:2015ira}:
\begin{widetext}
\begin{align}
  S_{\text{exact}}  & = \frac{N}{4\lalat} \sum_{\vn} \Tr{-\cFbar_{ab}(\vn)\cF_{ab}(\vn) - \chi_{ab}(\vn) \cD_{[a}^{(+)}\psi_{b]}^{\ }(\vn) - \eta(\vn) \cDbar_a^{(-)}\psi_a(\vn) + \frac{1}{2}\left(\cDbar_a^{(-)}\cU_a(\vn)\right)^2}, \\
  S_{\text{closed}} & = -\frac{N}{16\lalat} \sum_{\vn} \Tr{{\eps_{abcde}\ \chi_{de}(\vn + \hatbmu_a + \hatbmu_b + \hatbmu_c) \cDbar_c^{(-)} \chi_{ab}(\vn)}},
\end{align}
\end{widetext}
where \lalat is the dimensionless 't~Hooft coupling defined by $r_{\tau, \text{lattice}} = \lalat N_{\tau}^{4 - d}$.
The indices run from $1, \cdots, 5$, spanning the basis vectors of the $A_4^*$ lattice, and $\sum_{\vn}$ is over all lattice sites.
The $1 + 5 + 10$ fermion fields $\eta$, $\psi_a$ and $\chi_{ab} = -\chi_{ba}$ transform in representations of the $S_5$ point group symmetry, as do the five complexified gauge links $\cU_a$ and $\cUbar_a$ that combine the $4 + 6$ gauge and scalar field components.
These gauge links are used to form the complexified field strengths $\cF_{ab}$ and $\cFbar_{ab}$, as well as the finite difference operators $\cD_a^{(+)}$ and $\cDbar_a^{(-)}$.

In addition to these terms, we also include the two soft-supersymmetry-breaking deformations discussed in \secref{sec:lattice}.
$S_{\text{soft}}$ from \eq{eq:single_trace} is present to regulate flat directions even in four dimensions, while $S_{\text{center}}$ from \eq{eq:center} needs to be added once we specialize to the three-dimensional theory by setting $N_z = 1$.
The full three-dimensional lattice action is then
\begin{align}
  S_{\text{lattice}} & = S_{\text{exact}} + S_{\text{closed}} + S_{\text{soft}} + S_{\text{center}}.
\end{align}
As mentioned in \secref{sec:small}, we can omit $S_{\text{center}}$ (by setting its coefficient $\ka^2 = 0$) in the small-volume regime where the center symmetry in the reduced direction breaks dynamically.

Another detail mentioned in \secref{sec:small} is the need to remove the scalar-field contributions from the Wilson lines (spatial holonomies) that we analyze to distinguish between the small- and large-volume regimes.
As in \refcite{Catterall:2014vka, Catterall:2014vga, Catterall:2017lub}, we accomplish this by using a polar decomposition $\cU_a = H_a \cdot U_a$ to separate each $N\X N$ complexified gauge link into a positive-semidefinite hermitian matrix $H_a$ (containing the scalar fields) and a unitary matrix $U_a$ corresponding to the gauge field.
The resulting unitarized Wilson lines are simply the products $\prod_{i = 1}^{N_x} U_x(x_i, y, \tau)$ wrapping around the lattice, and similarly in the $y$-direction.
The distributions shown in Figs.~\ref{fig:hist_highT} and \ref{fig:hist_lowT} come from the Wilson lines in the $x$-direction, while the data released in \refcite{data} confirm that Wilson lines in both spatial directions are equivalent, as they should be for the $N_x = N_y$ we consider.

Since the lattice basis vectors are not orthogonal, in $d < 4$ dimensions the dimensionless lattice coupling \lalat has a non-trivial relation to the dimensionful continuum coupling $\la$.
Following the analysis in \refcite{Catterall:2017lub}, this relation can be written as
\begin{equation}
  r_{\tau, \text{lattice}} = \lalat N_{\tau} = \frac{(d+1)^{\frac{5-d}{8-2d}}}{\sqrt{d}} \la \be,
\end{equation}
which for three-dimensional SYM becomes
\begin{equation}
  \label{eq:couplings}
  \lalat N_{\tau} = \frac{4}{\sqrt{3}} \la \be.
\end{equation}
A standard quantity computed by our software is the dimensionless lattice bosonic action density $s_{\text{lat}}$ defined by~\cite{Catterall:2014vka, Schaich:2014pda}
\begin{equation}
  V(\Tbb^3) s_{\text{Bos}} = N_x N_y N_{\tau} \left(\frac{9N^2}{2}\right) (s_{\text{lat}} - 1),
\end{equation}
normalized and shifted in our conventions so that $s_{\text{lat}} = 1$ corresponds to unbroken supersymmetry.
Specializing to the aspect ratios $N_x / N_{\tau} = N_y / N_{\tau} = 1$ we consider in this work, we have
\begin{equation}
  \frac{N_x N_y N_{\tau}}{V(\Tbb^3)} = \frac{3\sqrt{3} N_{\tau}^3}{4\be^3} = 16 \frac{\la^3}{\lalat^3},
\end{equation}
using the relation between the couplings in \eq{eq:couplings}.
Plugging this in, we have
\begin{equation}
    -\frac{s_{\text{Bos}}}{N^2 \la^3} = \frac{72}{\lalat^3} (1 - s_{\text{lat}}).
\end{equation}
This expression connects the $s_{\text{lat}}$ data provided in \refcite{data} to the points shown in Figs.~\ref{fig:main}, \ref{fig:zeta_extrap}, and \ref{fig:cont_extrap}.
\end{appendix}

\raggedright
\bibliography{3d_2branes}
\end{document}